\documentstyle [12pt] {article}
\def\lsim{\lower2pt\hbox{$\buildrel{<}\over{\sim}$}}
\begin{document} 
\centerline{The Nexus between Cosmology and Elementary Particle Physics:}
\centerline{Testing Theoretical Speculations through Observations of the}
\centerline{Cosmic Microwave Background Anisotropies}
\vskip24pt 
\baselineskip=10pt 
\centerline{Mairi Sakellariadou}
\vskip12pt 
\centerline{D\'epartement de Physique Th\'eorique, 
              Universit\'e de Gen\`eve}
\centerline{24 quai Ernest-Ansermet, CH-1211 Gen\`eve 4, 
Switzerland}
\vskip 44pt 
\begin{abstract} 
\vskip 10pt 
The origin of the large scale structure in the universe --- galaxies, 
quasars, clusters, voids, sheets --- is one of the most important questions
in cosmology. 
One can show that some non-thermal energy density fluctuations must have 
been present in the early universe. These fluctuations grew by gravitational
instability to form the observed structures.  There are at present two
families of models to explain the origin of these initial fluctuations:
inflationary models and  topological defect scenarios.
Current observational developments provide a link with theoretical 
predictions, allowing us to test our theoretical models.
In this contribution, I present a sketch of the current status of 
the origin of cosmological structure formation.
\vspace{1cm} 
\end{abstract} 
\vspace{1cm} 

\section{Introduction}

At present, our understanding of the evolution of the observed universe 
rests on the hot big bang model, which, being particularly 
successful, is considered as the standard cosmology. In favor of this 
model, there is direct evidence which extends back to about $10^{-2}$ sec 
after the explosion, at the onset of primordial nucleosynthesis.
Combining theories of fundamental physics at ultra-high energies, with the
notion that standard cosmology is very robust, allows us to speculate 
about the history of our universe at times as early as $10^{-43}$ sec 
after the bang.

The standard cosmology, described by the Friedmann-Robertson-Walker metric,
lies upon three theoretical pillars: (i) the Einstein's general theory of 
relativity, which determines the dynamics of the universe; (ii) the 
cosmological principle, which states that the universe is homogeneous and 
isotropic on large scales; and (iii) a perfect fluid description of the 
matter content. On the other hand, its main observational pillars consist 
of: (i) the Hubble's redshift-distance relation, showing that the universe 
is expanding; (ii) the existence of a blackbody cosmic microwave 
background, discovered in 1965 by Penzias and Wilson \cite{pw}; and (iii) 
the agreement between observed and theoretically determined, according to 
nucleosynthesis, abundances of light elements. 

Despite its successes, the standard big bang model faces a number of 
unanswered questions, like the requirement up to a high degree of 
accuracy of an initially homogeneous and flat universe, the origin 
of the observed large scale structure, the small value of the
cosmological constant, the nature of the dark matter; as well as the problem 
of the age of the universe, a possible conflict between theory and 
observations. 

An appealing solution to the homogeneity and flatness problems is to 
introduce, during the very  early stages of the universe, a period of 
exponential expansion known as inflation \cite{guth}. 
According to the inflationary paradigm, the expansion of the universe was
driven, at an early stage of its history, by a scalar field.  During that
period the expansion was quite dramatic, and the quantum fluctuations of 
the scalar field \cite{qf} were enormously amplified when that phase ended.  
Thus, inflation provides a mechanism \cite{infl-pert} for the causal 
generation of the primordial density perturbations required for the observed 
large scale structures.

Inflationary models are not entirely free from problems and therefore it is
important to address the issue of the origin of structure formation according
to some other theory. The main alternative approach lies on the topological 
defect scenarios, based on the idea that a number of cosmological phase 
transitions took place, as the universe cooled down, associated with 
spontaneous symmetry breakings of the previous phase. Therefore, topological 
defects --- a well-studied phenomenon in condensed matter physics --- could 
have appeared in our universe \cite{kibble} and played the role of seeds for 
structure formation. Topological defect models have the advantage of 
depending on very few parameters and therefore are, in a sense, more appealing
than the inflationary ones. 
Depending on the nature of the broken symmetry, topological defects
can either be local or global, while their classification depends on the
number of components of an order parameter which breaks the symmetry
group.  Among the various topological defects, global monopoles, global
textures and both global or local cosmic strings, are viable candidates.
Since the initial density fluctuations have tiny amplitudes, their evolution
at early times can be studied using linear cosmological perturbation theory.

\section{Cosmic Microwave Background Radiation}

The Cosmic Microwave Background (CMB) radiation is the extraterrestrial
electromagnetic radiation that uniformly fills the space at wavelengths in the
range of millimiters to centimeters. 
At present, the spectrum of CMB radiation is, to a high degree of accuracy, a 
thermal Planck blackbody spectrum at a temperature \cite{mat} 
$T_0=2.728\pm 0.002 K$, as measured by the FIRAS (Far Infrared Absoulute 
Spectrophotometer) on the COBE (Cosmic Background Explorer) satellite 
developed by NASA. Since now the universe is optically thin to radio radiation,
the sea of CMB radiation, having almost completely relaxed to thermodynamic 
equilibrium, must be the remnant heat from the early hot and dense phase of 
the expanding universe. 

The existence of the CMB radiation with an almost thermal nature consists
the main evidence that the universe did indeed expand from a dense and hot 
state. In addition, the CMB radiation is extremeley close to isotropic;
this uniformity cannot be explained within the context of standard cosmology. 
The CMB radiation offers an essential probe of the origin of structure 
formation, through the effects of cosmological structure on the spectrum and 
isotropy of the relic CMB radiation. 

In 1992, the COBE-DMR (Differential Microwave Radiometer) experiment 
detected anisotropies (temperature irregularities) in the CMB radiation 
\cite{smoot}. These anisotropies, imprinted on the CMB radiation by primordial
perturbations generated within $10^{-35}$ sec after the big bang, were found to
be at the level $\Delta T/T \approx 1. \times 10^{-5}$ on all angular 
scales larger than $10^o$ and compatible with a scale invariant (spectral 
index of $n_s = 1.2 \pm 0.3$ \cite{gorski}) Harrison-Zel'dovich spectrum. 
While confirming the idea that indeed large structures grew from small 
initial fluctuations through gravitational instability, the COBE-DMR 
observations could not discriminate between inflationary models and topological
defect scenarios. The planned MAP (Microwave Anisotropy Probe) and 
COBRAS/SAMBA (COsmic Background Radiation Anisotropy Satellite / SAtellite for
Measurement of Background Anisotropies) satellite experiments, as well as 
ballons and ground based experiments, will probe anisotropies on smaller 
angular scales, which may allow us to distinguish between these two classes of
models for structure formation. Such experiments show how the early universe 
offers an ideal laboratory to test high energy physics models, on energy 
scales far beyond those of any conceivable terrestrial accelerator.
Moreover, from the point of view of a cosmologist, the CMB radiation offers
the unique way of determining basic cosmological parameters --- like $\Omega_0,
H_0, \Omega_b, \Lambda$ --- to within a few percent, through measurements of 
the CMB anisotropy spectrum.  The justification for this belief is mainly
that CMB anisotropies can be determined almost fully within linear cosmological
perturbation theory and are not particularly affected by nonlinear physics.

The CMB fluctuation spectrum is usually parametrized in terms of
multiple moments $C_\ell$, defined as the coefficients in the expansion of
the temperature autocorrelation function
\begin{equation}
 \langle{\delta T\over T}({\bf n}){\delta T\over T}({\bf n}')
\rangle\left|_{{~}_{\!\!({\bf n\cdot n}'=\cos\vartheta)}}\right. =
  {1\over 4\pi}\sum_\ell(2\ell+1)C_\ell P_\ell(\cos\vartheta)~, 
\end{equation}
which compares points in the sky separated by an angle $\vartheta$.  The value
of $C_\ell$ is determined by fluctuations on angular scales of order
$\pi/\ell$.  One usually plots $\ell(\ell+1)C_\ell$ versus $\ell$ --- known as
the power spectrum --- which is the power per logarithmic interval in $\ell$, 
giving the spectrum of anisotropies observed today.

For scalar perturbations, I will describe the main physical mechanisms which 
contribute to the redshift of photons propagating in a perturbed Friedmann 
geometry.
\hfill\break
(i) On large angular scales, the main contribution to CMB anisotropies comes 
from inhomogeneities in the spacetime geometry. These inhomogeneities 
determine the change in the photon energy, due to the difference of the 
gravitational potential at the position of emitter and observer, and account 
for red-shifting or blue-shifting, caused by the time dependence of the 
gravitational field along the path of the photon. They are known as 
``ordinary'' Sachs-Wolfe and Integrated Sachs-Wolfe (ISW) effects respectively.
\hfill\break
(ii) On angular scales 
$0.1^\circ\stackrel{<}{\sim}  \theta\stackrel{<}{\sim}  2^\circ$, the main 
contribution comes from the intrinsic inhomogeneities on the surface of the 
last scattering, due to acoustic oscillations in the coupled baryon-radiation 
fluid prior to decoupling.  On the same  angular scales  as this acoustic 
term, there is a Doppler contribution to the CMB anisotropies, due to the 
relative motions of emitter and observer. The sum of these two contributions 
is denoted by the term ``acoustic peaks''.
\hfill\break
(iii) On scales smaller than about $0.1^{\circ}$,  the anisotropies
are damped due to the finite thickness of the recombination shell,
as well as by photon diffusion during recombination (Silk damping).

Both, generic inflationary models and topological defect scenarios, predict an
approximately scale-invariant spectrum of density perturbations on large
angular scales ($\ell\ \lsim \ 50$), thus the COBE-DMR data provide mainly a
normalization for the different models.  Cosmic microwave background
anisotropies on intermediate and small angular scales are very important.
If the two classes of theories predict different characteristics for the 
acoustic peaks (e.g., amplitude and position of primary peak, existence or 
absence of secondary peaks) we can discriminate among them.  In the nearby 
future, a number of sophisticated experiments will scrutinize various regions 
of the sky trying to reveal the characteristics of the relic CMB radiation.

\section{Families of Models for Structure Formation}

Within the framework of gravitational instability theory, there are two
currently investigated families of models to explain the origin of the
observed structure in the universe.
\hfill\break  
(i) Initial density perturbations can be due to freezing in of quantum 
fluctuations of a scalar field during an inflationary era
\cite{stein}. Such  fluctuations were produced at a very early time in the
history of the universe, and were driven far beyond the Hubble radius by the 
enormous (inflationary) expansion.  As a result, inflationary fluctuations
are not altered anymore and evolve freely according to homogeneous linear
perturbation equations until basically the time of galaxy formation.
Moreover, as a result of the nature of quantum fluctuations,
the distribution of amplitudes of these initial perturbations is usually
Gaussian. 
\hfill\break  
(ii) Initial density perturbations can be seeded by an  inhomogeneously 
distributed form of energy, called ``seed'', which contributes only a small 
fraction to the total energy density of the universe and which interacts with 
the cosmic fluid only gravitationally. A familiar example is the case of
topological defects, which could have appeared naturally during a symmetry 
breaking phase transition in the early universe \cite{kibble}.
According to these models, cosmological structure was formed as a result
of a symmetry breaking phase transition and a phase ordering.
Such initial fluctuations are  generated continuously and evolve according to 
non-homogeneous linear perturbation equations.
Perturbations from defect models are generally non-Gaussian.

On large angular scales, both families of models 
predict an approximately scale-invariant Harrison-Zel'dovich spectrum
\cite{h, z}, the Sachs-Wolfe plateau.
The acoustic peaks on intermediate scales in the CMB power spectrum,
might represent a mean to support or rule out one of these two families of 
models. 

In the case of inflationary models, there has been a large number of studies
and a lot of excitement, in particular since CMB anisotropies might lead to a 
determination of fundamental cosmological parameters, such as the spatial 
curvature of the universe $\Omega_0$, the baryon density $\Omega_b$, the 
Hubble constant $H_0$ and the cosmological constant $\Lambda$.
At multipoles $\ell \ge 200$, the CMB anisotropies become sensitive to 
fluctuations inside the Hubble horizon at recombination. Since these
fluctuations had enough time to evolve prior to last scattering, they are 
sensitive to evolutionary effects that depend on a number of cosmological 
parameters. The new generation of satellites (MAP and especially 
COBRAS/SAMBA) having high sensitivity, angular resolution and large sky 
coverage, are expected to provide a mean to determine these fundamental 
cosmological parameters to a precision of a few percent. 

The power spectrum predicted for a generic inflationary model reveal the 
existence of a primary peak at $\ell\sim 200$ with an amplitude $\sim (4-6)$
times the Sachs-Wolfe plateau, and the existence of secondary oscillations
\cite{stein}.

On the other hand, seed models (like topological defect models), generally 
predict a quite different power spectrum than inflation, due to the
behaviour of perturbations on super-horizon scales.  Causality and 
scale invariance have quite different implementations in this class of models.
While in inflationary models randomness appears only when initial conditions
are set up and the time evolution is linear and deterministic, in seed models
randomness also appears during the time evolution, as a result of a complex
non-linear process.  Seed models are more complicated to be solved than 
inflationary ones, due to the fact that the linear perturbation equations are
non-homogeneous with a source term due to the seed. Since the seed evolution 
is, in general, a non-linear and complicated process, much less precise
predictions have been made so far, and there is a limited number of studies on
the family of seed models.

Recent studies \cite{ct, ram, joao} on generic topological 
defect models show that the primary acoustic peak is located to the right of
the adiabatic position, at which the peak arises in a generic inflationary 
model. The value of this shift to smaller angular scales is determined by 
the coherence length of the defect. Also the structure of secondary peaks 
may be quite different for generic defects as compared to inflation.  
Depending on whether the defect is effectively coherent or not, which is a 
direct implication of the constraints  imposed by causality on defect 
formation and evolution, secondary peaks will or will not appear in the 
power spectrum \cite{joao}.  Considering density perturbations seeded by 
global textures, $\pi_3$ defects \cite{turok}, in a universe dominated by 
cold dark matter,
the position of the primary acoustic peak was found to be displaced by
$\Delta \ell\sim 150$  towards smaller angular scales than in standard 
inflationary models \cite{ct, ram}, while its amplitude was only a 
factor of $\sim 1.5 -3.3 $ times higher than the Sachs-Wolfe plateau 
\cite{ram}.  In an attempt to reveal the robust features of the power
spectrum in a seed model, it was found \cite{nt, new} that there are defect 
models leading to a primary acoustic peak located at the adiabatic position, 
however its amplitude may be susbtantially smaller than the one in generic
inflationary models \cite{new}.

The satellites MAP and, in particular, COBRAS/SAMBA 
with planned launch years around 2000 and 2005 are designed to image 
anisotropies of the CMB radiation to an uncertainty better than 
$\Delta T/T \sim 2 \times 10^{-6}$ at all angular scales larger than $\sim$ 
10 arcmin over the whole sky.  We therefore expect to have a full power
spectrum against which we could test our theoretical models.

\section{Conclusions}

The nexus between cosmology and elementary particle physics has become an  
especially active area of research in recent years. Current frontiers of  
particle physics involve energy scales far beyond those available now or in  
the near future terrestrial particle accelerators.  An obvious place to look 
is to the very early universe, where conditions of extreme energy and density 
are realized. At the same time, the standard big bang model provides a  
reliable framework for describing the evolution of our universe as early as  
$10^{-2}$ sec after the explosion, when the temperature was about $10$ MeV.   
Extending our understanding to earlier times and higher temperatures, requires
knowledge about the fundamental particles and their interactions at very high  
energies; progress in cosmology has become linked to progress in particle  
physics. 

Among the main, still open problems in modern cosmology, remains the origin 
of the observed structure in the universe.  Based on all present indications,
we believe that the large-scale structure was produced by gravitational 
instability from small primordial fluctuations in the energy density, 
generated during the early stages of the universe. Within this framework,
the two families of models to explain the origin of  primordial density 
perturbations are inflationary models and topological defect scenarios.
Either of these two families of models predicts precise 
fingerprints in the cosmic microwave background anisotropies, which can
be used to differentiate among these models using a purely linear 
analysis. Both families lead to approximately scale-invariant 
Harrison-Zel'dovich spectrum of density fluctuations on large angular scales.
However, the power spectrum predicted from each of these families, has 
different properties on smaller angular scales. The next satellite
experiments (MAP and COBRAS/SAMBA), as well as ground-based (e.g., Jodrell 
Bank, CAT, SASKATOON, VSA) and ballons experiments (e.g., BOOMER, FIRS, MAX, 
MAXIMA, MSAM, UCSB) will provide a detailed power spectrum, against which
we will be able to test our theoretical models.  In addition, we expect to be
able to determine a number of fundamental cosmological parameters up to a high 
degree of accuracy.  It is therefore believed that these coming years will
be particularly fruitful for cosmology and one may conclude that as
cosmologists we are currently living through what can be considered a 
scientific revolution.

\bigskip
\bigskip
\hfill\break
{\bf Acknowledgements}

\hfill\break
It is a pleasure to thank the members of the Jet Propulsion Laboratory,
who organized this workshop and, in particular, Lute Maleki, for inviting
me to deliver this talk.

\end{document}